\documentclass[aps,showpacs,prl,twocolumn]{revtex4}
\usepackage{graphicx}

\begin{document}

\title{Probing the nodal gap in the pressure-induced heavy fermion superconductor CeRhIn$_5$}
\author{Tuson Park$^{1,2}$, E. D. Bauer$^1$, and J. D. Thompson$^1$}
\affiliation{$^1$Los Alamos National Laboratory, Los Alamos, New Mexico 87545, USA\\ $^2$ Department of Physics, Sungkyunkwan University, Suwon 440-746, Korea}
\date{\today}

\begin{abstract}
We report field-orientation specific heat studies of the pressure-induced heavy fermion superconductor CeRhIn$_5$. Theses experiments provide the momentum-dependent superconducting gap function for the first time in any pressure-induced superconductor. In the coexisting phase of superconductivity and antiferromagnetism, field rotation within the Ce-In plane reveals four-fold modulation in the density of states, which favors a d-wave order parameter and constrains a theory of the interplay between superconductivity and magnetism.
\end{abstract}
\maketitle

The emergence of superconductivity near an antiferromagnetic quantum critical point (QCP) suggests that magnetic fluctuations associated with the QCP may provide a glue to form  electron pairs~\cite{mathur98}. In unconventional superconductors, the superconducting (SC) order parameter, which is tied closely to the SC pairing mechanism, breaks the underlying crystalline symmetry and contains nodes at which the SC gap becomes zero on the Fermi surface. The presence of nodal quasiparticles qualitatively changes the density of states (DOS) at the Fermi level and leads to a power-law temperature dependence of thermodynamic properties~\cite{volovik93}. This power-law behavior contrasts with the exponential temperature dependence in fully gapped conventional superconductors and, therefore, provides an indicator of unconventional superconductivity. Unambiguous determination of an unconventional SC order parameter, however, comes from phase-sensitive and momentum-dependent
studies that, complemented by the thermodynamic measurements, have established a d-wave order parameter in high-$T_c$ cuprates~\cite{harlingen95} and  p-wave symmetry in Sr$_2$RuO$_4$ ~\cite{kidwingira06}.

In heavy fermion superconductors, where the relationship between superconductivity and quantum criticality is most conspicuous~\cite{mathur98}, details of the SC gap are rarely available because of disorder from chemical substitution or pressure environments that are needed to induce superconductivity. In this Letter, we report field-orientation specific heat ($C_p$) measurements on the pressure-induced superconductor CeRhIn$_5$. When a magnetic field is rotated in the Ce-In plane, four-fold modulation in $C_p$ is observed deep in the SC state, with minima along [100] that indicate gap zeros at this direction in a d-wave order parameter. These results provide the momentum-dependent SC gap function for the first time, which is also the first in any pressure-induced superconductor, and open a new avenue to insights on the pairing mechanism in strongly correlated superconductors.

CeRhIn$_5$ is an antiferromagnet below $T_N$ = 3.8~K and belongs to a family of 
heavy-fermion compounds CeMIn$_5$ (M=Co, Ir, Rh) that become superconducting for M=Co and Ir at 2.3 and 0.4~K, respectively~\cite{hegger00, petrovic01, roman01}. Derived from the antiferromagnet CeIn$_3$, it crystallizes in the tetragonal HoCoGa$_5$ structure, where a Ce-In plane and a M-In block are alternately stacked along the c-axis. When subjected to pressure, CeRhIn$_5$ becomes superconducting near 0.5~GPa where $T_N$ starts to decrease, suggesting a correlation between superconductivity and magnetism (see Fig.~1a)~\cite{hegger00, tuson06}. As shown in Fig.~1a, there are two critical pressures at 1.75 and 2.3~GPa denoted P1 and P2. Below P1, superconductivity and magnetism coexists on a microscopic scale, while a purely SC phase only appears above P1~\cite{mito03}. P2 is a field-tuned quantum critical point hidden by the SC dome~\cite{tuson06}. At pressures above P1, the zero-field nuclear spin-relaxation rate $1/T_1$ and specific heat show a power-law temperature dependence below $T_c$, which is similar to what is found in the ambient-pressure heavy-fermion superconductor CeCoIn$_5$ with a d-wave SC order parameter~\cite{mito03, fisher02, park08}. In the coexisting phase below P1, the temperature dependence of $1/T_1$ crosses over from $T^3$ to $T$-linear with decreasing temperature in the SC state, suggesting a change from  d-wave singlet pairing to an exotic SC order parameter~\cite{kawasaki03}. Singular quantum fluctuations in the vicinity of an AFM QCP also have been proposed to favor unconventional order parameters ranging from a p-wave singlet to extended d-wave with additional nodal points~\cite{fuseya03, bang04}.
\begin{figure}[tbp]
\centering  \includegraphics[width=7.0cm,clip]{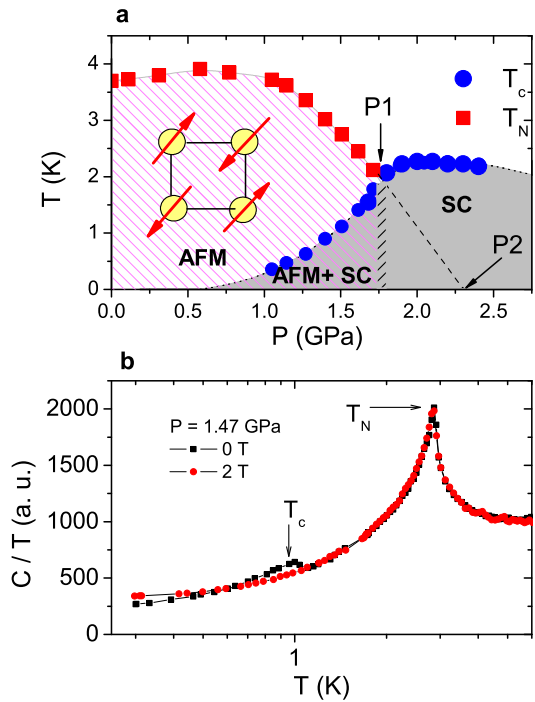}
\caption{(color online) \textbf{a} Phase diagram of CeRhIn$_5$ as a function of pressure~\cite{tuson08}. Pressure induces a bulk superconducting (SC) state at 0.5~GPa that coexists with the antiferromagnetic (AFM) phase up to P1, where $T_c$ becomes equal to $T_N$. The critical pressure marked by P2 is the field-tuned quantum critial point hidden by the SC dome. \textbf{b} Heat capacity of CeRhIn$_5$ at 1.47 GPa. The incommensurate AFM transition at 2.86~K is followed by a SC transtion at 1.04~K. When 2~T is applied (circles), the magnetic transition is barely changed, while the superconductivity is completely destroyed.}
\label{figure1}
\end{figure}

Whereas power laws in $1/T_1$ and $C_p$ are consistent with a nodal gap in the coexistence phase, these measurements are insensitive to the location of nodes. To resolve this important question, we have developed a field-rotation specific heat technique under pressure. Single crystals used in this study were obtained from a batch where a residual resistivity ratio (RRR) was measured routinely to be of the order of 500, indicating exceptional sample quality. A hybrid clamp-type pressure cell made of Be-Cu/NiCrAl and silicone-fluid transmitting medium were used to produce hydrostatic pressure environments up to 2.5~GPa. The superconducting transition temperature of Sn was  measured inductively to determine the pressure at low temperatures. Magnetic field rotation was performed by using a triple-axis vector magnet with a large bore of $2.5"$ (American Magnetic Inc.) that accommodated a $^3$He cryostat with the pressure cell. Specific heat under pressure was measured by an ac calorimetric technique where ac heating incurs oscillation in the sample temperature ($T_{ac}$) that is inversely proportional to the sample's heat capacity, i.e., $T_{ac}\propto 1/C$~\cite{sullivan68}. The oscillating temperature was measured by a Au-Fe/chromel thermocouple, which was attached to one face of the plate-like sample, while a constantan heater was glued to the opposite face. Magnetic field effects on the thermoelectric voltage of the Au-Fe thermocouple were determined independently.

Figure 1b shows the temperature dependence of the heat capacity of CeRhIn$_5$ in its coexisting phase at 1.47~GPa. Long-range magnetic order with wave vector \textbf{Q}=(0.5, 0.5, 0.297) appears at 2.86~K ($T_N$) and is robust against magnetic field; whereas, the SC transition occurs at 1.04~K ($T_c$) and is completely destroyed at 2~T ($H\parallel$~a-axis). The magnetic transition at this pressure is as sharp as it is at ambient pressure, indicating a negligibly small pressure gradient in the pressure cell. 

\begin{figure}[tbp]
\centering  \includegraphics[width=7.0cm,clip]{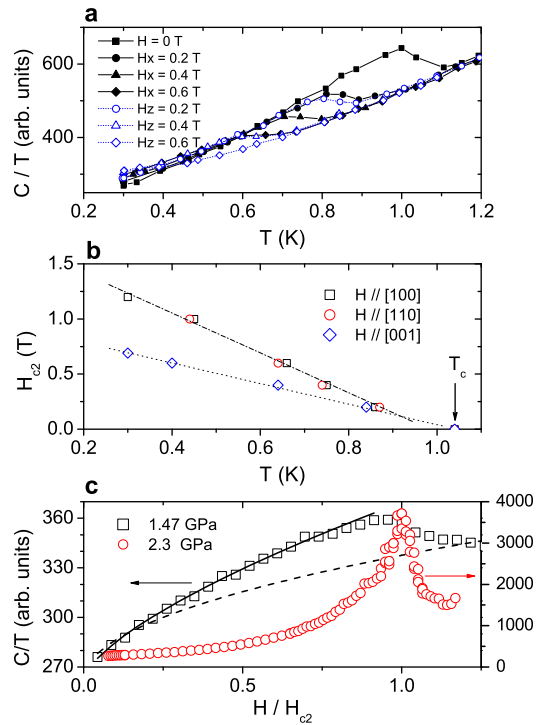}
\caption{(color online) \textbf{a} Low-temperature heat capacity ($C_p$) of CeRhIn$_5$ at 1.47~GPa and in low applied fields indicated in the legend. Solid and open symbols represent $C_P/T$ for field along [100] and [001], respectively. \textbf{b} Temperature dependence of the upper critical field $H_{c2}$. Squares, circles, and diamond symbols correspond to fields along [100], [110], and [001], respectively. The dotted (or dash-dotted) line is a guide for a linear-$H$ dependence. \textbf{c} Magnetic field dependence of $C/T$ at 300~mK for $H \parallel$[100] at 1.47~GPa (squares) and 2.3~GPa(circles). Here $H_{c2}$ is 1.2 and 7.7~T for 1.4 and 2.3~GPa, respectively. The solid line describes least-squares fitting by $\Delta C/T = C/T-C(0T)/T=AH^n$ with $n=0.68$, while the dashed line is for $n=0.5$.}
\label{figure2}
\end{figure}
The low-$T$ heat capacity of CeRhIn$_5$ is shown in Fig.~2a for representative fields along [100] (solid symbols) and along [001] (open symbols). These data allow a determination of anisotropy of the upper critical field $H_{c2}$. Within the ab-plane, the upper critical field $H_{c2}^{ab}$ is essentially isotropic (see Fig.~2b), but $H_{c2}^c$ along the c-axis is smaller by a factor of two, which reflects the tetragonal crystalline structure as found in CeCoIn$_5$~\cite{settai01}. For magnetic field along any crystalline axis, $H_{c2}$ reveals a linear-$T$ dependence down to the lowest experimental temperature (0.3~K), which is different from CeCoIn$_5$ where Pauli paramagnetic effects dominate conventional orbital fields $(H_{c2}^{orb} =-0.73T_c~dH_{c2}/dT)$ and $H_{c2}$ is smaller than $H_{c2}^{orb}$ by a factor of 5~\cite{bianchi03}. Considering the exceptional sample quality of CeRhIn$_5$ ($\rho_0 = 30~n\Omega \cdot cm$), the lack of Pauli limiting may arise from the presence of AFM that allows a SC order parameter other than a spin singlet~\cite{sigrist91}. We also note a positive curvature in $H_{c2}^{ab}$ near $T_c$, which may be ascribed to multi-band effects since there are three major bands that contribute to charge conductivity in CeRhIn$_5$~\cite{shishido05}.

Field dependence of the specific heat of CeRhIn$_5$ does not conform to expectations for conventional type-II superconductors. Figure~2c shows representative field-swept data at 300~mK for H along [100] at 1.47~GPa (squares) and at 2.3~GPa (circles). In fully gapped conventional superconductors, $C/T$, which is proportional to the density of states (DOS), linearly depends on magnetic field because contribution to the DOS comes from quasiparticles generated inside vortex cores and the number of vortices is proportional to $H$. In unconventional superconductors with nodes, however, $C/T$ increases as $\sqrt{H}$ because Doppler-shifted quasiparticles leak through the nodes and contribute to the DOS~\cite{ichioka99}. A power-law form of $C/T =C(0T)/T+AH^n$ was used to fit the field-dependent specific heat of CeRhIn$_5$ at 1.47~GPa and the solid line in Fig.~2c best describes the data with $n=0.68\pm0.04$ for field along [100]. This anomalous behavior may reflect the interplay between superconductivity and coexisting antiferromagnetism~\cite{tuson08, cornelius01}. The field dependence of $C/T$ at 2.3~GPa is strikingly different from that in the coexisting phase, showing a first-order like sharp feature at $H_{c2}$. The upper critical field of CeCoIn$_5$ also is first order below $T_c/2$, suggesting that CeCoIn$_5$ is similar to the high-pressure phase of CeRhIn$_5$~\cite{tuson06,ikeda01,aoki04}.  

\begin{figure}[tbp]
\centering  \includegraphics[width=6.5cm,clip]{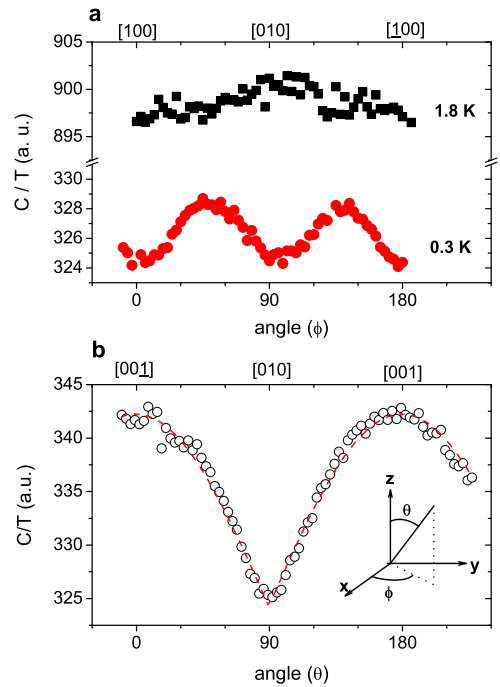}
\caption{(color online) Field-rotation specific heat of CeRhIn$_5$ at 1.47~GPa and 300~mK. \textbf{a} Azimuthal rotation ($\phi$) of the magnetic field (0.5~T) within the Ce-In plane both in the AFM (1.8~K) and the coexisting phase of AFM and SC (0.3~K). \textbf{b} Polar rotation ($\theta$) of the magnetic field (0.5~T) perpendicular to the Ce-In plane in the coexisting phase at 0.3~K. The solid line is a least-squares fit to a two-fold modulation model: $C=C_0+A(1+A_2|cos\phi|)$ with $C_0=265$, $A=59.3$, and $A_2=0.30$, where the oscillation arises from $H_{c2}$ anisotropy.}
\label{figure3}
\end{figure}
Figure~3 shows the field-angle specific heat of CeRhIn$_5$ at 1.47~GPa for field rotations $\phi$ ($\theta$) within (perpendicular to) the Ce-In plane. In the paramagnetic state at 3~K, $C/T$ is featureless as a function of angle (not shown). In the AFM state but above $T_c$ (squares in Fig.~3a), $C/T$ shows a broad feature around [010] that may reflect the presence of AFM domains below $T_N$ (=2.86~K). Upon decreasing temperature below $T_c$ (=1.04~K), a new modulation in $C(\phi)/T$ appears in the mixed SC state (circles in Fig.~3a). The four-fold modulation is distinctly different from the spectrum observed in the magnetic state, indicating that the new structure is not related to magnetism, but arises from the SC gap symmetry. A polar sweep at the same pressure, temperature and field (Fig. ~3b) shows only a two-fold modulation that reflects the factor of two anisotropy in $H_{c2}$ between the ab-plane and c-axis. 

Field-rotation specific heat, which is proportional to the angle-dependent DOS at low temperature, probes the momentum-dependent SC gap function and identifies gap-zero directions in unconventional superconductors~\cite{vekhter99, tuson03, miranovic05, sakakibara07}. In the Abrikosov state, supercurrent circulates around a vortex core and extended quasiparticles experience a Doppler energy shift, $\delta E \approx \bf{v_s} \cdot \bf{k_F}$, where $\bf{k_F}$ is the Fermi momentum of the quasiparticles and $\bf{v_s}$ is the supercurrent velocity. Derived from the Doppler effects that depend on the relative magnetic field orientation against nodal points on the Fermi surface, the DOS precisely reflects the anisotropic gap and modulates with field-rotation angle such that the DOS is minimal for field along nodal points but is maximal for field along antinodes~\cite{vekhter99}. Contributions to the field-rotation specific heat can be written as $C_{total}(\phi)=C_0+C_2(\phi)+C_4(\phi)$, where the constant $C_0$ is the zero-field specific heat arising from thermally excited quasiparticles and phonons, the two-fold $C_{2}(\phi)$ comes from sample misalignment, and $C_4 (\phi)$ describes modulation from the SC order parameter. The field-angle $\phi$ was determined against the crystalline a-axis or [100]. The lack of a two-fold oscillation indicates that the ab-plane is well matched with the field-rotation plane. Figure~4a is a plot of the field-induced heat capacity divided by temperature $\Delta C/T$ at 0.3~K and 1.47~GPa as a function of magnetic field angle, where $\Delta C=C_{total}-C_0=C_4(\phi)$. Minima in $\Delta C$ can be approximated by a parabolic form $A(1-A_4 cos4\phi)$ (dashed line) that is expected for a model gap structure of $\Delta (\theta,\phi)=\Delta_0 \sqrt{1+cos4\phi}$ on a three-dimensional Fermi surface. Here $\theta$ is a polar angle determined against the c-axis or [001].

\begin{figure}[tbp]
\centering  \includegraphics[width=7.0cm,clip]{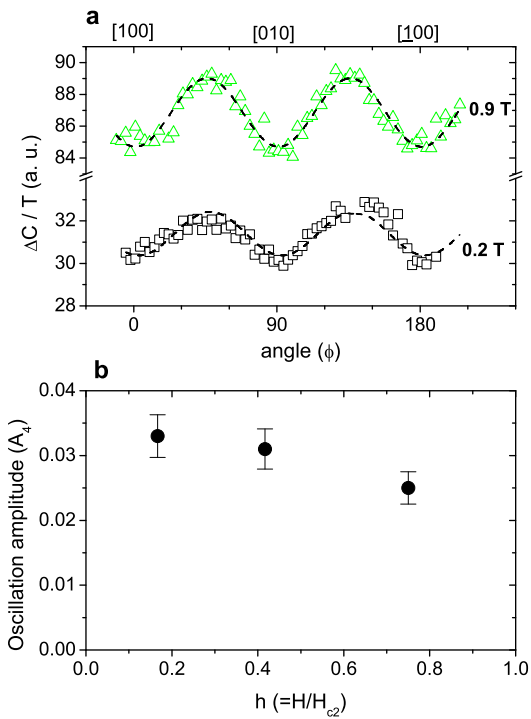}
\caption{(color online) \textbf{a} Field-rotation specific heat of CeRhIn$_5$ at 0.3~K for 0.2 (squares) and 0.9~T (triangles). Dashed lines are least-square fits by a sinusoidal form of the angular dependence (see text). \textbf{b} Oscillation amplitude $A_4$ as a function of the normalized field $h (=H/H_{c2})$ with $H_{c2}=1.2~T$ at 0.3~K.}
\label{figure4}
\end{figure}
Figure 4b shows the four-fold oscillation amplitude $A_4$ at 0.3~K and 1.47~GPa as a function of magnetic field, which is obtained by a least squares fit to the sinusoidal form (dashed line in Fig.~4a). Compared to negligible $H_{c2}$ anisotropy within the ab-plane (Fig.~2b), the large oscillation amplitude ($\approx 3.3\%$) in $\Delta C(\phi)$ of CeRhIn$_5$ indicates that the four-fold modulation is not due to Fermi surface anisotropy, but due to symmetry of the SC gap function~\cite{miranovic05, sakakibara07}. Like CeCoIn$_5$, minima in $\Delta C$ of CeRhIn$_5$ are located along $<100>$, indicating $d_{xy}$ symmetry with line node along that direction~\cite{sakakibara07}. In a recent theoretical treatment, however, Vorontsov and Vekhter argued that a reversal in the DOS anisotropy occurs at a critical field due to additional pair breaking from vortex scattering; consequently, minima along $<100>$ in $C_p$ can be interpreted instead as evidence for a $d_{x^2-y^2}$ gap~\cite{vorontsov06}. A key future experiment to distinguish the two d-wave order parameters is to identify the critical field at which anisotropy reversal occurs in the field-rotation specific heat of CeRhIn$_5$.

These results constrain possible candidates of the SC gap in the coexisting phase of antiferromagnetism and superconductivity of CeRhIn$_5$. Four-fold modulation of the in-plane specific heat excludes an extended d-wave with additional nodes at points where the antiferromagnetic Brillouin zone crosses the Fermi surface, which was suggested to account for the $T$-linear $1/T_1$ at lowest temperatures~\cite{bang04}. A gapless p-wave spin-singlet state that was predicted~\cite{fuseya03} to be favored over a d-wave symmetry near a QCP is also inconsistent with our observations. The two-fold modulation in specific heat under polar field sweeps, as shown in Fig.~3b, rules out the possibility of a hybrid gap with nodes at the poles, which has been suggested for CeIrIn$_5$~\cite{shakeripour07}. Instead, these results indicate a d-wave symmetry with line nodes along the c-axis.

To summarize, four-fold and two-fold modulations in representative in-plane and out-of-plane field-rotation specific heat measurements on CeRhIn$_5$ favor a d-wave SC order parameter over other suggested candidates in the vicinity of the quantum critical point, thus constraining a theory of the interplay between superconductivity and antiferromagnetism. This work will assure further studies of other pressure-induced superconductors where the nature of the SC order parameter has yet to be identified.

The authors thank I. Vekhter for discussion. Work at Los Alamos was performed under the auspices of the U.S. Department of Energy/Office of Science and supported by the Los Alamos LDRD program. T.P acknowledges a grant from the Korea Science and Engineering Foundation (KOSEF) funded by the Korea government R01-2008-000-10570-0.

\end{document}